\documentclass{article}

\usepackage{graphicx}
\usepackage{subfigure}
\usepackage{url}
\usepackage{here}

\usepackage{color} 
\definecolor{darkgreen}{rgb}{0,0.5,0}
\usepackage[colorlinks=true,%
  linkcolor=red,%
  citecolor=darkgreen,%
  urlcolor=blue]{hyperref}

\usepackage{fullpage}

\usepackage{soul}
\usepackage{xspace}

\newcommand{\Am}{\ensuremath{\mathbf{A}}}
\newcommand{\Xm}{\ensuremath{\mathbf{X}}}
\newcommand{\Ym}{\ensuremath{\mathbf{Y}}}

\newcommand{\xv}{\ensuremath{\mathbf{x}}}
\newcommand{\yv}{\ensuremath{\mathbf{y}}}

\begin{document}

\title{Performance Evaluation of Sparse Matrix Multiplication Kernels
  on Intel Xeon Phi}

\author{
  Erik Saule$^\dag$, Kamer Kaya$^\dag$, and \"{U}mit V. \c{C}ataly\"{u}rek$^{\dag\ddag}$\\
  $\dag$ The Ohio State University, Department of Biomedical Informatics\\
  $\ddag$ The Ohio State University, Department of Electrical and Computer Engineering\\
  email:{\it \{esaule,kamer,umit\}@bmi.osu.edu}
}

\maketitle
\begin{abstract}
Intel Xeon Phi is a recently released high-performance coprocessor
which features $61$ cores each supporting $4$ hardware threads with
$512$-bit wide SIMD registers achieving a peak theoretical performance
of 1Tflop/s in double precision. Many scientific applications involve
operations on large sparse matrices such as linear solvers,
eigensolver, and graph mining algorithms. The core of most of these
applications involves the multiplication of a large, sparse matrix
with a dense vector~(SpMV). In this paper, we investigate the
performance of the Xeon Phi coprocessor for SpMV. We first provide a
comprehensive introduction to this new architecture and analyze its
peak performance with a number of micro benchmarks.
Although the design of a
Xeon Phi core is not much different than those of the cores in modern
processors, its large number of cores and hyperthreading capability
allow many application to saturate the available memory bandwidth,
which is not the case for many cutting-edge processors. Yet, our
performance studies show that it is the memory latency not the
bandwidth which creates a bottleneck for SpMV on this architecture.
Finally, our experiments show that Xeon Phi's sparse kernel
performance is very promising and even better than that of
cutting-edge general purpose processors and GPUs.\looseness=-1

\end{abstract}

\section{Introduction}
Given a large, sparse, $m \times n$ matrix $\Am$, an input vector
$\xv$, and a cutting edge shared-memory multi/many core architecture
Intel Xeon$\textregistered$ Phi, we are interested in analyzing the
performance obtained while computing $$\yv \gets \Am\xv$$ in
parallel. The computation, known as the sparse-matrix vector
multiplication~(SpMV), and some of its variants, such as the
sparse-matrix matrix multiplication~(SpMM), form the computational
core of many applications involving linear systems, eigenvalues, and
linear programs, i.e., most large scale scientific
applications. For this reason, they have been extremely intriguing in
the context of high performance computing~(HPC). Efficient
shared-memory parallelization of these kernels is a well studied area
~\cite{Bell09,Buluc2009_SPAA,Buluc11,Krotkiewski10,Williams07}, and
there exist several techniques such as prefetching, loop
transformations, vectorization, register, cache, TLB blocking, and
NUMA optimization, which have been extensively investigated to
optimize the
performance~\cite{Im01,Mellor-Crummey04,Nishtala07,Williams07}. In
addition, company-built support is available for many
parallel shared-memory architectures, such as Intel's MKL and NVIDIA's
cuSPARSE. Popular 3rd party sparse matrix libraries such as
OSKI~\cite{Vuduc05} and pOSKI~\cite{Jain08} also exist.\looseness=-1

Intel Xeon Phi is a new coprocessor with many cores,
hardware threading capabilities, and wide vector registers. The
novel architecture is of interest for HPC community: the Texas Advanced
Computing Center has built the Stampede cluster which is equipped with
Xeon Phi and many other computing centers are following. Furthermore,
some effort has been pushed to have a fast MPI
implementation~\cite{Potluri12-TACC}. Although Intel Xeon Phi has
been released recently, performance evaluations already exist in
literature~\cite{Eisenlohr12-TACC,Saule12-MTAAP,Stock12-TACC}.
Eisenlohr~et~al. investigated the behavior of dense linear algebra
factorization on Xeon Phi~\cite{Eisenlohr12-TACC} and
Stock~et~al. proposed an automatic code optimization approach for
tensor contraction kernels~\cite{Stock12-TACC}. Both of these works
work on dense and regular data. in a previous work, two of the authors evaluates the scalability
of graph algorithms, coloring and breadth first search~(BFS)~\cite{Saule12-MTAAP}. None of these works give absolute performance
values and to the best of our knowledge there exist no such work in
the literature. Although similar to BFS, SpMV
and SpMM are different kernels~(in terms of synchronization, memory
access, and load balancing), and as we will show, the new coprocessor
is very promising and can even perform better than existing
cutting-edge CPUs and accelerators while handling sparse linear
algebra.\looseness=-1

Accelerators/coprocessors are designed for specific tasks. They do not
only achieve a good performance but also reduce the energy usage per
computation. Up to now, GPUs have been successful w.r.t. these
criteria and they reported to perform well especially for regular
computations. However, the irregularity and sparsity of SpMV-like
kernels create several problems for these architectures. In this
paper, we analyze how Xeon Phi performs on two popular sparse linear
algebra kernels, SpMV and SpMM. \cite{cramer2012openmp} studied the
performance of a Conjugate Gradient application which uses SpMV,
however this study concerns only a single matrix and is application
oriented. To the best of our knowledge, we give the first analysis of
the performance of the coprocessor on these kernels.\looseness=-1

We conduct several experiments with 22 matrices from UFL Sparse Matrix
Collection\footnote{\url{http://www.cise.ufl.edu/research/sparse/matrices/}}.
To have architectural baselines, we also measured the performance of
two dual Intel Xeon processors, X5680~(Westmere) E5-2670~(Sandy
Bridge), and two NVIDIA Tesla$\textregistered$ GPUs C2050 and K20, on
these matrices. Overall, Xeon Phi's sparse matrix performance is very
promising and even better than that of general purpose CPUs and the GPUs we used
in our experiments. Furthermore, its large registers and extensive
vector instruction support boost its performance and make it a
candidate for both scientific computing and throughput oriented
server-side code for SpMV/SpMM-based services such as product/friend
recommendation.\looseness=-1

Having 61 cores and hyperthreading capability can help the Intel Xeon
Phi to saturate the memory bandwidth during SpMV, which is not the
case for many cutting edge processors.  Yet, our analysis showed
that the memory latency, not the memory bandwidth, is the bottleneck and
the reason for not reaching to the peak performance. As expected, we
observed that the performance of the SpMV kernel highly depends on the
nonzero pattern of the matrix and its sparsity: when the nonzeros in a
row are aligned and packed in cachelines in the memory, the
performance would be much better. We investigate two different
approaches of densifying the computation either by ordering the matrix
via the well known reverse Cuthill-McKee ordering~{\em
RCM}~\cite{Cuthill69} or changing the nature of the computation by
 register blocking.\looseness=-1

The paper is organized as follows: Section~\ref{sec:mic} presents a
brief architectural overview of the Intel Xeon Phi coprocessor along
with some detailed read and write memory bandwidth micro
benchmarks. Section~\ref{sec:ker} describes the sparse-matrix
multiplication kernels and Sections~\ref{sec:spmv} and~\ref{sec:spmm}
analyze Xeon Phi's performance on these kernels. The performance
comparison against other cutting-edge architectures is given in
Section~\ref{sec:comp}. Section~\ref{sec:con} concludes the paper.\looseness=-1

\section{The Intel Xeon Phi Coprocessor}
\label{sec:mic}

In this work, we use a pre-release KNC card SE10P. The card has 8
memory controllers where each of them can execute 5.5 billion
transactions per second and has two 32-bit channels. That is the
architecture can achieve a total bandwidth of 352GB/s aggregated
across all the memory controllers. There are 61 cores clocked at
1.05GHz. The cores' memory interface are 32-bit wide with two channels
and the total bandwidth is 8.4GB/s per core. Thus, the cores should be
able to consume 512.4GB/s at most. However, the bandwidth between the
cores and the memory controllers is limited by the ring network which
connects them and theoretically supports at most 220GB/s.\looseness=-1

Each core in the architecture has a 32kB L1 data cache, a 32kB L1
instruction cache, and a 512kB L2 cache. The architecture of a core is
based on the Pentium architecture: though its design has been updated
to 64-bit. A core can hold 4 hardware contexts at any time. And at
each clock cycle, instructions from a single thread are executed. Due
to the hardware constraints and to overlap latency, a core never
executes two instructions from the same hardware context
consecutively. In other words, if a program only uses one thread, half
of the clock cycles are wasted. Since there are 4 hardware contexts
available, the instructions from a single thread are executed
in-order. As in the Pentium architecture, a core has two different
concurrent instruction pipelines~(called U-pipe and V-pipe) which
allow the execution of two instructions per cycle. However, some
instructions are not available on both pipelines: only one vector
or floating point instruction can be executed at each cycle, but two
ALU instructions can be executed in the same cycle.\looseness=-1

Most of the performance of the architecture comes from the vector processing
unit. Each of Intel Xeon Phi's cores has 32$\times$512-bit SIMD
registers which can be used for double or single precision, that is, either
as a vector of 8$\times$64-bit 
values or as a vector of 16$\times$32-bit values, respectively. The
vector processing unit
can perform many basic instructions, such as addition or division, and
mathematical operations, such as sine and sqrt, allowing to reach 8
double precision operations per cycle~(16 single precision). The unit
also supports {\it Fused Multiply-Add}~(FMA) operations which are
typically counted as two operations for benchmarking
purposes. Therefore, the peak performance of the SE10P card is 1.0248
Tflop/s in double precision~(2.0496 Tflop/s in single precision) and
half without FMA.\looseness=-1

\subsection{Read-bandwidth benchmarks}
\label{sec:read-bw}

\begin{figure*}[htb]
\center
\subfigure[Simple sum, {\tt char}, {\tt -O1}]{\includegraphics[width=.49\linewidth,page=3]{data/simplesum-char-O1.pdf}\label{fig:read-8bit}}
\subfigure[Simple sum, {\tt int}, {\tt -O1}]{\includegraphics[width=.49\linewidth,page=4]{data/simplesum-int-O1.pdf}\label{fig:read-32bit}}\\
\subfigure[Vectorization]{\includegraphics[width=.49\linewidth,page=1]{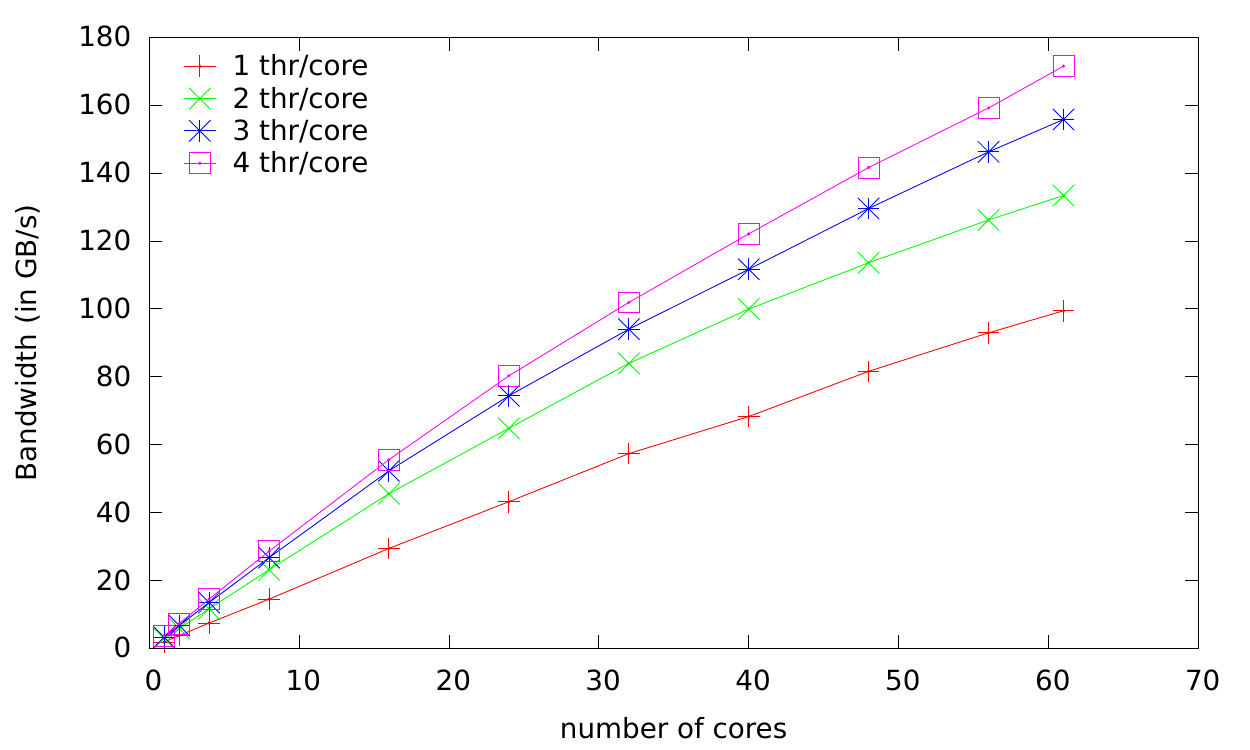}\label{fig:read-vect}}
\subfigure[Vectorization + prefetching]{\includegraphics[width=.49\linewidth,page=2]{data/vectsum-pref.pdf}\label{fig:read-vect-pref}}
\caption{\baselineskip=2pt\small{The read bandwidth obtained by Intel Xeon Phi on our
  micro-benchmarks containing different array sum operations. In (a)
  and (b), an upper bound on bandwidth is computed by the 1.05Ghz core
  speed where 4 instructions are required per {\tt int} and 5
  instructions~(including an additional casting) is required per {\tt
  char}. In (c) and (d), the theoretical upper bound on maximum
  bandwidth, $\max$(8.4GB/s $\times$ \#cores, 220GB/s), due to maximum
  per core and ring bandwidths is shown.\looseness=-1}}
\label{fig:sum-perf}
\end{figure*}

We first try to estimate what is the maximum achievable read bandwidth
in Xeon Phi. By benchmarking the performance on simple operations by
different number of cores and threads per core, we can have an idea
about where the performance bottlenecks come from. To that end, we
choose the array {\it sum kernel} which just computes the sum of large
arrays. And we measure the amount of effective (application) bandwidth
reached. To compute the sum, each core first allocates 16MB of data
and then each thread reads the whole data. To minimize cache reusage,
each thread starts from a different array. All the results of this
experiment are presented in Figure~\ref{fig:sum-perf} where the
aggregated bandwidth is reported for {\tt char} and {\tt int} data
types.\looseness=-1

First we sum the array as a set of 8-bit integers, i.e., {\tt char}
data type. The benchmark is compiled via {\tt -O1} optimization flag
so as to prevent the compiler optimizations with possible undesired
effects. The results are shown in Figure~\ref{fig:read-8bit}.  The
bandwidth peaks at 12GB/s when using 2 threads per core and 61 cores.
The bandwidth scales linearly with the number of cores indicating that
the benchmark does not saturate the memory bus. Moreover, additional
threads per core do not bring significant performance improvements
which indicates a full usage of the computational resources in the
cores. To show that we have plotted peak effective bandwidth one can expect to
achieve if instructions can utilize both pipelines (denoted as {\em Full
  Pairing} in the figure) or not (denoted as {\em No Pairing}). For this
particular code, compiled executable have 5 instructions per loop. As seen in
the figure, best achieved performance aligns well with No Pairing line,
indicating that those 5 instructions were not paired, and this benchmark is
instruction bound. \looseness=-1

The second benchmark performs the summation on the arrays containing
32-bit {\tt int}s. Again, we employ the {\tt -O1} optimization
flag. The results are presented in Figure~\ref{fig:read-32bit}. This
code has 4 instructions in the loop and process 4 times more data in each
instruction, hence the performance is roughly scaled 5 times. The
performance still scales linearly with the number of cores indicating
the performance is not bounded by the speed of the memory
interconnect. The peak bandwidth, 60.0GB/s, is achieved by using 4
threads per core. There is only a small difference with the bandwidth
achieved using 3 threads (59.9GB/s), which implies that the benchmark
is still instruction bound. Though we can notice that 2 threads were
not enough to achieve peak bandwidth (54.4GB/s) and at least 3 threads
are necessary to hide memory latency.\looseness=-1

The third benchmark also considers the array as a set of 32-bit
integers but uses vector operations to load and sum 16 of them at a
time, i.e., it processes 512 bits, a full cacheline, at once.  The
results are presented in Figure~\ref{fig:read-vect}. The bandwidth
peaks at 171GB/s when 61 cores and 4 threads per core are used. This
benchmark is more memory intensive than the previous ones, and even 3
threads per core can not hide memory latency. Moreover, although it is
not very clear in the figure, the bandwidth scales sub-linearly with
the increasing number of cores. This indicates the bandwidth is
getting closer to the speed of the actual memory bus.\looseness=-1

The fourth benchmark tries to improve on the previous one by removing
the necessity of hiding the memory latency via using memory
prefetching. The results are presented in
Figure~\ref{fig:read-vect-pref}. The bandwidth peaks at 183GB/s when 61
cores and 2 threads per core are used (representing 3GB/s per core). The bandwidth with a single
thread per core is relatively worse (149GB/s) and scales linearly with
the number of cores. However, with 2 threads, the performance no
longer scales linearly from 24 cores and seems to have almost
plateaued at 61 cores. A single core can sustain 4.8GB/s of read bandwidth when it is
alone accessing the memory.\looseness=-1

\subsection{Write-bandwidth benchmarks}
\label{sec:write-bw}

\begin{figure*}[htb]
\subfigure[Store]{\includegraphics[width=.33\linewidth]{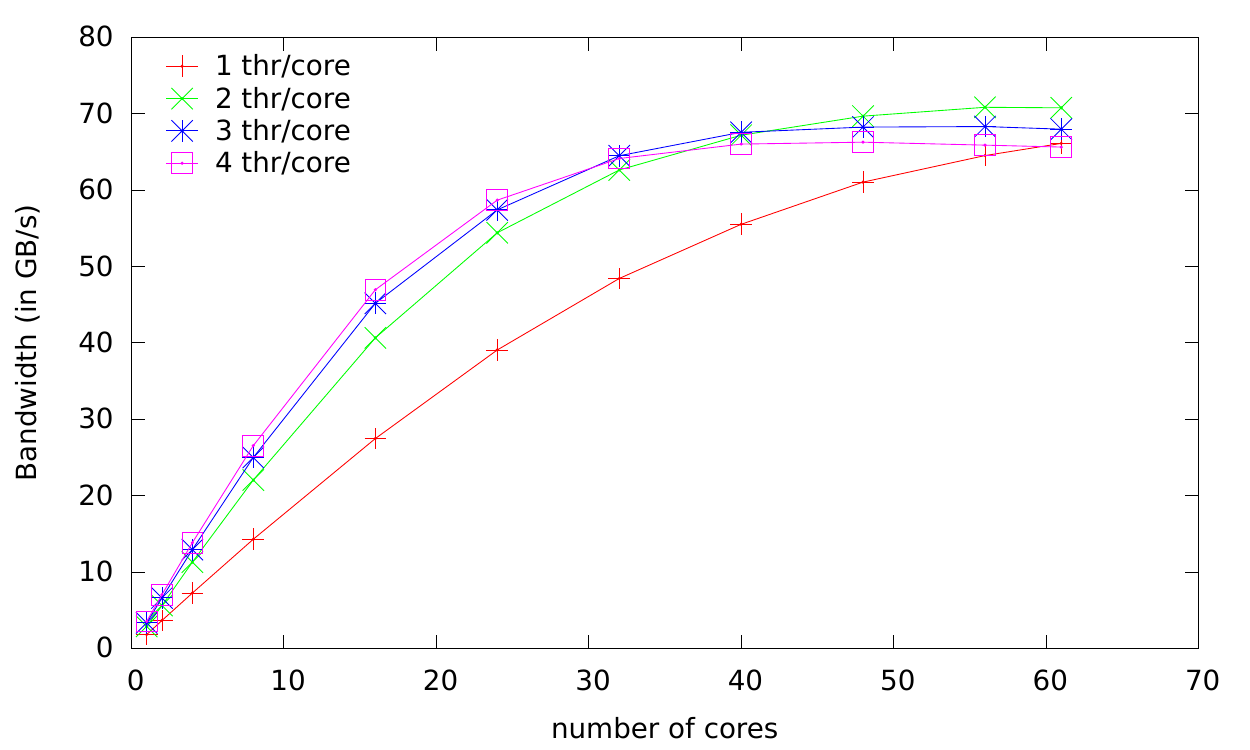}\label{fig:write-store}}
\subfigure[Vectorization + No-Read hint]{\includegraphics[width=.33\linewidth]{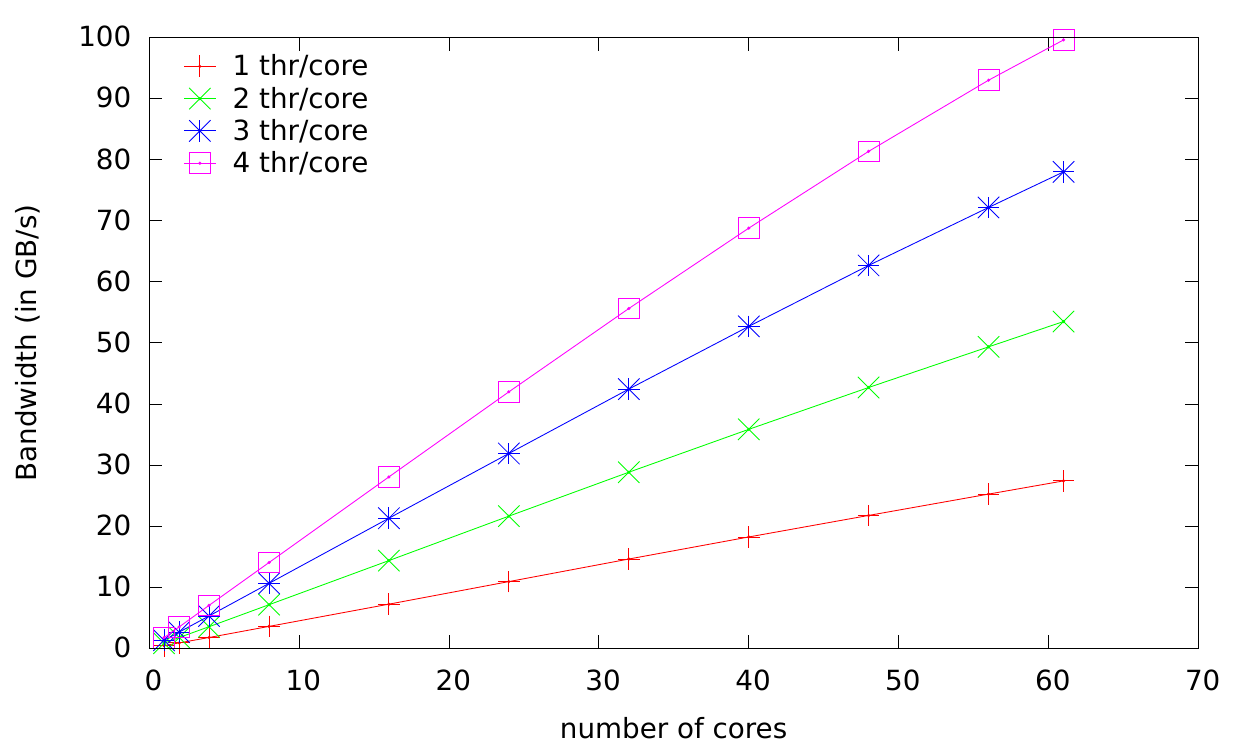}\label{fig:write-nr}}
\subfigure[Vectorization + No-Read hint + No global ordering]{\includegraphics[width=.33\linewidth,page=2]{data/memset-nrngo.pdf}\label{fig:write-nrngo}}
\caption{The write bandwidth obtained by Intel Xeon Phi on three
  micro-benchmarks using different approaches for {\it memsetting} the
  buffer. In (c), the theoretical upper bound on maximum bandwidth due
  to maximum per core and ring bandwidths is shown.}
\label{fig:write-mem}
\end{figure*}

After the read bandwidth, we also want to investigate how write
bandwidth behaves with a set of micro benchmarks in which each thread
allocates a local 16MB buffer and fills it with some predetermined
value. The results are presented in Figure~\ref{fig:write-mem}.\looseness=-1

From the previous micro benchmarks on read bandwidth, it should be
clear that non-vectorized loops will not perform well since they will
be instruction bound, so we do not consider non-vectorized cases. The
first benchmark uses fills the memory by crafting a buffer in a
512-bit SIMD register and writing a full cache line at once using the
simple store instruction. As shown in Figure~\ref{fig:write-store},
when all 61 cores are used, a bandwidth of 65-70GB/s is achieved for
each thread per core configuration.  The second benchmark writes to
memory using the {\it No-Read hint}.  This bypasses
the {\it Read For Ownership} protocol which forces to read a cacheline
even before overwriting its whole content. Note that the protocol is
important in case of partial cacheline updates. The results of that
benchmark are presented in Figure~\ref{fig:write-nr}. The achieved
bandwidth scales linearly up to 100GB/s using 61 cores and 4 threads
per core. It also improves linearly with the number of threads which
indicates that the cores are stalling write operations.\looseness=-1

The third benchmark uses a different instruction to perform the memory
write. Regular write operations enforces that the writes are committed
to the memory in the same order the instructions were emitted. This
property is crucial to ensure the coherency in many shared memory
parallel algorithms. The instruction used in this benchmark allow the
writes to be {\it Non-Globally Ordered}. (This also uses the No-Read
hint). The results of the benchmark are presented in
Figure~\ref{fig:write-nrngo}. With this instruction, the 100GB/s
bandwidth obtained in the previous experiment is achieved with only 24
cores. The bandwidth peaks at 160GB/s when all 61 cores are used
(representing 2.6GB/s per core). Note
that it can be achieved by using a single thread per core. This
instruction allows not to stall the threads until the internal buffer
of the architecture is full, and a single thread can manage to fill it
up.A single core can sustain 5.6GB/s of write bandwidth when it is
alone accessing the memory.\looseness=-1

\section{Sparse Multiplication Kernels}\label{sec:ker}

\begin{figure}[htbp]
\centering
\subfigure[SpMV]{\includegraphics[width=.32\linewidth]{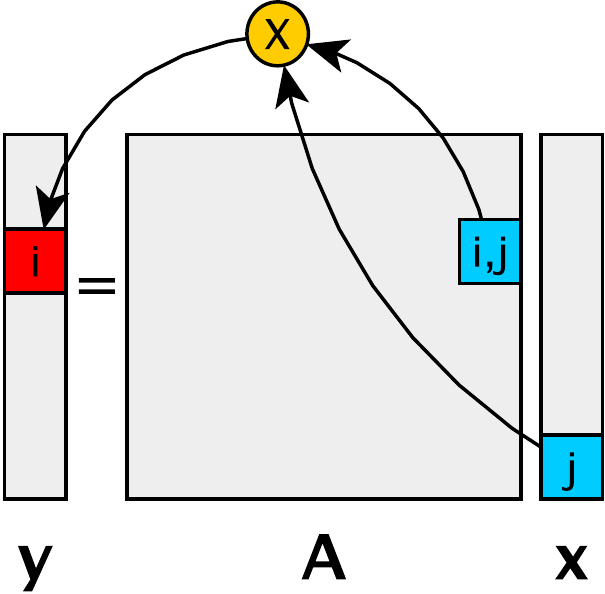}\label{fig:spmv}}\hspace*{5ex}
\subfigure[SpMM]{\includegraphics[width=.39\linewidth]{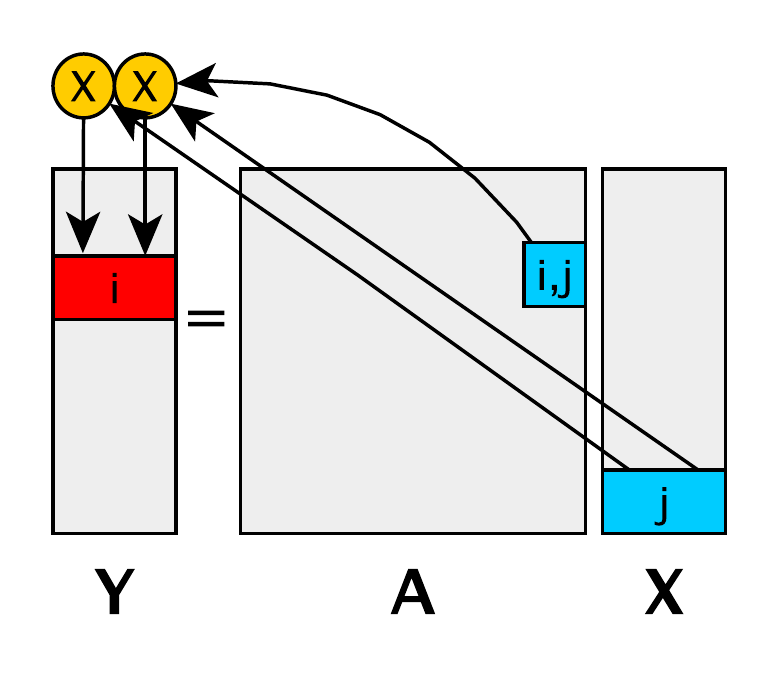}\label{fig:spmm}}
\caption{Memory access patterns for SpMV and SpMM. The color blue shows reads
  and red shows writes.}
\label{fig:spmvm}
\end{figure}

The most well-known sparse-matrix kernel is in the form $\yv\gets
\Am\xv$ where $\Am$ is an $m \times n$ sparse matrix, and $\xv$ and
$\yv$ are $n \times 1$ and $m \times 1$ column vectors~(SpMV). In this
kernel, each nonzero is accessed, multiplied with an $\xv$-vector
entry, and the result is added to a $\yv$-vector entry once. That is,
there are two reads and one read-and-write per each nonzero
accessed. Another widely-used sparse multiplication kernel is the
sparse matrix-dense matrix multiplication~(SpMM) in which $\xv$ and
$\yv$ are $n \times k$ and $m \times k$ dense matrices. In SpMM, there
are $k$ reads and $k$ read-and-writes per each nonzero accessed. The
read and write access patterns of these kernels are given in
Figure~\ref{fig:spmv}.
Obtaining good performance for SpMV is difficult almost on any
architecture. The difficulty arises from  the sparsity
pattern of $\Am$ which yields a non-regular access to the memory. The amount of
computation per nonzero is also very small. Most of the
operations suffer from bandwidth limitation.\looseness=-1

An $m \times n$ sparse matrix $\Am$ with $\tau$ nonzeros is usually
stored in the compressed row (or column) storage format~CRS (or
CCS). Two formats are dual of each other. Throughout the paper, we
will use CRS hence, we will describe it here. CRS uses three arrays:\looseness=-1
\begin{itemize}\itemsep -2pt
\item{{\it cids}[.] is an integer array of size $\tau$ that stores the column ids
  for each nonzero in row-major ordering.}
\item{{\it rptrs}[.] is an integer array of size $m+1$. For $0 \leq i <
  m$, {\it rptrs}[i] is the location of the first nonzero of $i$th row in the {\it
  cids} array. The first element is {\it rptrs}[0] $= 0$, and the last
  element is {\it rptrs[m]} $= \tau$. Hence, all the column indices of row
  $i$ is stored between {\it cids}[{\it rptrs}[i]] and ${\it
  cids}[{\it rptrs}[i+1]]-1$.
\item{\it val}[.] is an array of size $\tau$. For $0 \leq i < \tau$,
  {\it val}[i] keeps the value corresponding to the nonzero in {\it cids}[i]}.
\end{itemize}

There exist other sparse matrix representations~\cite{Saad94sparskit}, and the best storage format almost
always depends on the pattern of the matrix and the
kernel. In this work, we use CRS as it constitutes a
solid baseline.  Since $\Am$ is represented in CRS, it is a
straightforward task to assign a row to a single thread in a parallel
execution. As Figure~\ref{fig:spmv} shows, each entry $\yv_i$ of the
output vector can be computed independently while streaming the matrix
row by row. While processing a row $i$, multiple $\xv$ values are
read, and the sum of the multiplications is written to $\yv_i$. Hence,
there are one multiply and one add operation per nonzero, and the
total number of floating point operations is $2\tau$.\looseness=-1

\section{SpMV on Intel Xeon Phi}\label{sec:spmv}

For the experiments, we use a set of $22$ matrices in total whose
properties are given in Table~\ref{tab:matrices}. The matrices we are
taken from the UFL Sparse Matrix Collection with one exception {\it
mesh\_2048} which corresponds to a 5-point stencil mesh in 2D with
size $2048 \times 2048$.  We store the matrices in memory by using the
CRS representation. All scalar values are stored in double
precision~(64-bit floating point). And all the indices are stored via
$32$-bit integers. For all the experiments and figures in this
section, the matrices are numbered from $1$ to $22$ by increasing
number of nonzero entries, as listed in the table. For all the experiments, we first run the
operation 70 times and compute the averages of the last 60 operations
which are shown in figures and tables. Caches are flushed between each
measurement.\looseness=-1

\renewcommand{\tabcolsep}{0.08cm}
\begin{table}
\caption{Properties of the matrices used in the experiments. All
  matrices are square.}
\centering
\scalebox{0.90}{
\begin{tabular}{llrrlrrr}
   &      &       &    &        &        & max  & max\\
\# & name & \#row & \#nonzero& density& nnz/row& nnz/r& nnz/c
\\\hline
1 & {\it shallow\_water1} & 81,920 & 204,800 & 3.05e-05  & 2.50 & 4 & 4\\
2 & {\it 2cubes\_sphere}  & 101,492 & 874,378 & 8.48e-05 & 8.61 & 24 & 29\\
3 & {\it scircuit} & 170,998 & 958,936 & 3.27e-05 & 5.60  & 353 & 353\\
4 & {\it mac\_econ}  & 206,500 & 1,273,389 & 2.98e-05  & 6.16 & 44 & 47\\
5 & {\it cop20k\_A} & 121,192  & 1,362,087 & 9.27e-05 &  11.23 & 24 & 75\\
6 & {\it cant} & 62,451  & 2,034,917 & 5.21e-04 & 32.58 & 40 & 40\\
7 & {\it pdb1HYS} & 36,417  & 2,190,591 & 1.65e-03 &  60.15 & 184 & 162\\
8 & {\it webbase-1M}  & 1,000,005 & 3,105,536 & 3.10e-06  & 3.10 & 4700 & 28685\\
9 & {\it hood}  & 220,542 & 5,057,982 & 1.03e-04 & 22.93 & 51 & 77\\
10& {\it bmw3\_2}  & 227,362 & 5,757,996 & 1.11e-04 & 25.32 & 204 & 327\\
11& {\it pre2} & 659,033 & 5,834,044 & 1.34e-05 &  8.85 & 627 & 745\\
12& {\it pwtk} & 217,918 & 5,871,175 & 1.23e-04 &  26.94 & 180 & 90\\
13& {\it crankseg\_2}  & 63,838 & 7,106,348 & 1.74e-03 & 111.31 & 297 & 3423\\
14& {\it torso1} & 116,158 & 8,516,500 & 6.31e-04 & 73.31 & 3263 & 1224\\
15& {\it atmosmodd} & 1,270,432 & 8,814,880 & 5.46e-06 & 6.93 & 7 & 7  \\
16& {\it msdoor}  & 415,863 & 9,794,513 & 5.66e-05 &  23.55 & 57 & 77  \\
17& {\it F1} & 343,791 & 13,590,452 & 1.14e-04  & 39.53 & 306 & 378\\
18& {\it nd24k} & 72,000 & 14,393,817 & 2.77e-03 & 199.91 & 481 & 483\\
19& {\it inline\_1} & 503,712 & 18,659,941 & 7.35e-05 & 37.04 & 843 & 333\\
20& {\it mesh\_2048}  & 4,194,304 & 20,963,328 & 1.19e-06 & 4.99 & 5 & 5 \\
21& {\it ldoor} & 952,203 & 21,723,010 & 2.39e-05 & 22.81 & 49 & 77\\
22& {\it cage14} & 1,505,785 & 27,130,349 & 1.19e-05 & 18.01 & 41 & 41
\end{tabular}
\label{tab:matrices}}
\end{table}

\subsection{Performance evaluation}

\begin{figure}[htb]
\includegraphics[width=.96\linewidth]{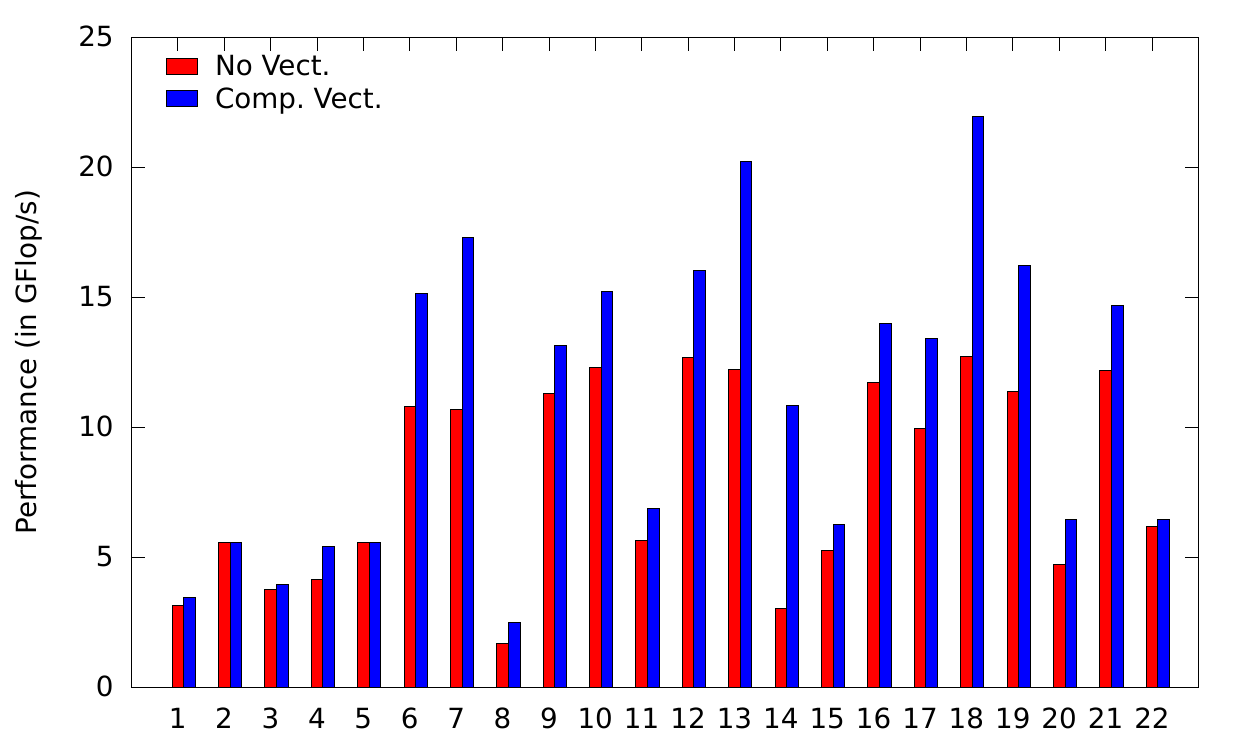}
\caption{Comparing {\tt -O1}~(No Vect.) and {\tt -O3}~(Comp. Vect.)
for SpMV in Xeon Phi.}
\label{fig:res-spmv-o1o3}
\end{figure}

The SpMV kernel is implemented in C++ using OpenMP and processes the
rows in parallel. We tested our dataset with multiple scheduling
policies. Figure~\ref{fig:res-spmv-o1o3} presents the highest
performance obtained when compiled with {\tt -O1} and {\tt -O3}. In
most cases, the best performance is obtained either using 61 cores and
3 threads per core, or using 60 cores and 4 threads per core. The
dynamic scheduling policy with a chunk size of 32 or 64 is
typically giving the best results.\looseness=-1

When compiled with {\tt -O1}, the performance obtained varies from $1$
to $13$GFlop/s. Notice that the difference of performance is not
correlated to the size of the matrix. When compiled with {\tt -O3} the
performance rises for all matrices and reaches 22GFlop/s on {\it
nd24k}. In total, $5$ matrices from our set achieve a performance over
15GFlop/s.\looseness=-1 

An interesting observation is that the difference on performance is
not constant; it depends on the matrix. Analyzing the assembly code
generated by the compiler for the SpMV inner loop~(which computes the
dot product between a sparse row of the matrix and the dense input
vector) gives an insight on why the performances differ. When {\tt
-O1} is used, the dot product is implemented in a simple way: the
column index of the nonzero is put in a register. And the value of the
nonzero is put in another. Then, the nonzero is multiplied with the
input vector element~(it does not need to be brought to register, a
memory indirection can be used here). The result is accumulated. The
nonzero index is incremented and tested for looping purposes. In other
words, the dot product is computed one element at a time with 3 memory
indirections, one increment, one addition, one multiplication, one
test, and one jump per nonzero.\looseness=-1

The code generated in {\tt -O3} is much more complex. Basically, it
uses vectorial operations so as to make 8
operations at once. 
The compiled code loads 8 consecutive values of the
sparse row in a 512-bit SIMD register~(8 double precision floating
point numbers) in a single operation. Then it populates another SIMD
register with the values of the input vector. Once populated, the two
vectors are multiplied and accumulated with previous results in a
single Fused Multiply-Add operation. Populating the SIMD register with
the appropriate values of the $\xv$ is non trivial since these
values are not consecutive in memory. One simple~(and
inefficient) method would be to fetch them the one after the
other. However, Xeon Phi offers an instruction, {\it
vgatherd}, that allows to fetch multiple values at once. The
instruction takes an offset array~(in a SIMD register), a pointer to
the beginning of the array, a destination register, and a bit-mask
indicating which elements of the offset array should be loaded. In the
general case, the instruction can not load all the elements at once, it
can only simultaneously fetch the elements that are on the same
cacheline. The instruction needs to be called as many times as the
number of cachelines the offset register touches. So overall, one FMA,
two vector loads (one for the nonzero from the matrix and one for the
column positions), one increment, one test, and some vgatherd are
performed for each 8 nonzeros of the matrix.\looseness=-1

\begin{figure}[htb]
\includegraphics[width=.96\linewidth,page=1]{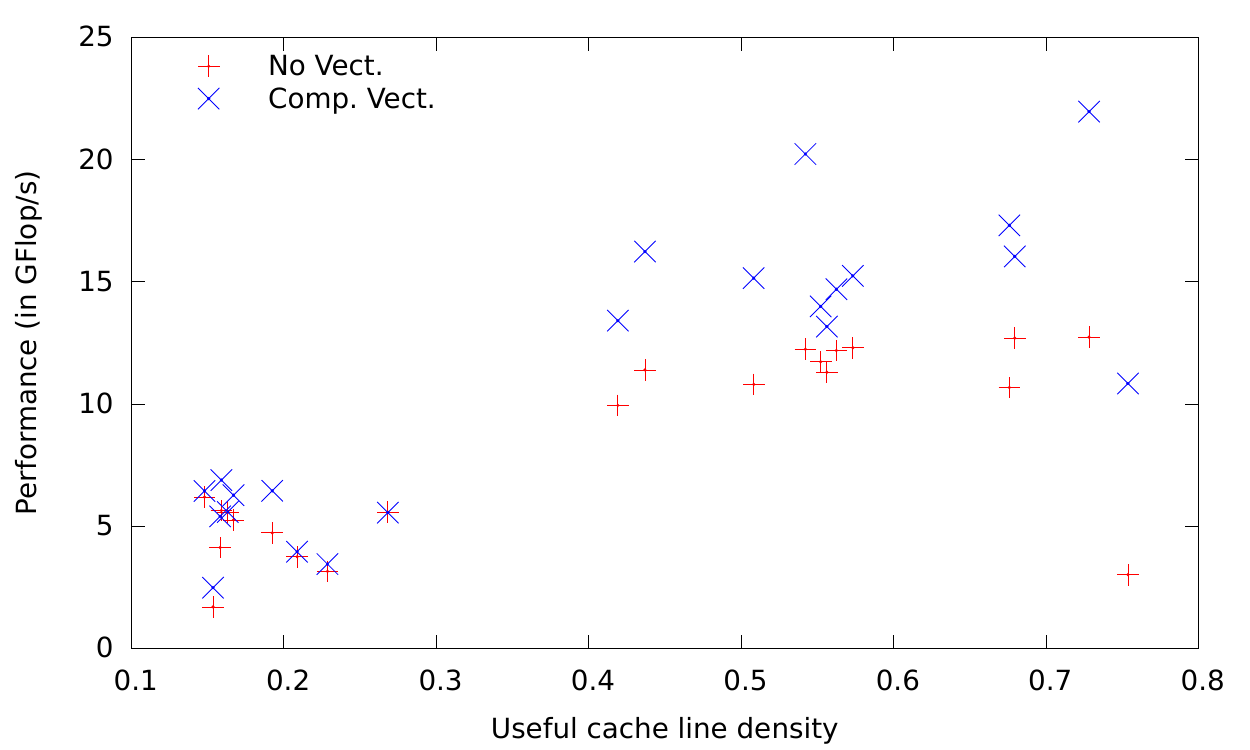}
\caption{The improvement of {\tt
        -O3} (Comp. Vect.) is linked to cacheline density.}
\label{fig:spmv-cld}
\end{figure}

Figure~\ref{fig:spmv-cld} shows the SpMV performance for each matrix
with {\tt -O1} and {\tt -O3} flags as a function of the {\it useful
cacheline density}~(UCLD), a metric we devised for the analysis. For
each row, we computed the ratio of the number of nonzeros on that row
to the number of elements in the cachelines of the input vector due to
that row. Then we took the average of these values to compute UCLD.
For instance, if a row has three nonzeros, 0, 19 and 20, the UCLD for
that single row would be $\frac{3}{16}$ since there are three nonzeros
spread on two cachelines of the input vector~(the one storing 0-7 and
the one storing 18-25). That is the worst possible value for UCLD is
$\frac{1}{8} = 0.125$ when only one element of each accessed cacheline
is used and the best possible value is $1$ when the nonzeros are
packed into well aligned regions of $8$ columns. For each matrix of
our data set, there are two points in Figure~\ref{fig:spmv-cld}, one
giving the performance in {\tt -O1}~(marked with `+'s) and one giving
the performance in {\tt -O3}~(marked with `$\times$'s). These two
points are horizontally aligned for the same matrix, and the vertical distance between them
represents the corresponding improvement on the performance for the
matrix. The performance improvement with
vectorization, and in particular with vgatherd, is significantly much
higher when the UCLD is high. Hence, the maximum performance achieved
with vectorial instructions is fairly correlated with UCLD.
However, this correlation does not explain what is bounding the
computation. The two most common source of performance bound are
memory bandwidth and floating point instruction throughput. For Intel
Xeon Phi, the peak floating point performance of the architecture is
about 1TFlop/s in double precision. Clearly, that is not the
bottleneck for our experiments.\looseness=-1

\subsection{Bandwidth considerations}

The nonzeros in the matrix need to be transferred to the core before
being processed. Assuming the access to the vectors do not incur any
memory transfer, and since each nonzero takes 12 bytes~(8 for the
value and 4 for the column index) and incurs two floating point
operations~(multiplication and sum), the flop-to-byte ratio of SpMV is
$\frac{2}{12} = \frac{1}{6}$. For our micro benchmarks, we saw that
the sustained memory bandwidth is about 180GB/s, which indicates a maximum
performance for the SpMV kernel of 30GFlop/s.\looseness=-1

Assuming only 12 bytes per nonzero need to be transfered to the core
gives only a {\it naive bandwidth} for SpMV: both vectors and the row
indices also need to be transferred. For an $n \times n$ matrix with
$\tau$ nonzeros, the actual minimum amount of memory that need to be
transferred in both ways is $2\times n \times 8 + (n+1)\times 4 + \tau
\times (8+4) = 4 + 20 \times n+ 12\times \tau$. If the density of the
matrix is high $12\tau$ will dominate the equation, but for sparser
matrices, $20n$ should not be ignored. The {\it application
bandwidth}, which tries to take both terms into account, is a common
alternative cross-architecture measure of performance on SpMV.\looseness=-1

\begin{figure}[htb]
\includegraphics[width=.95\linewidth]{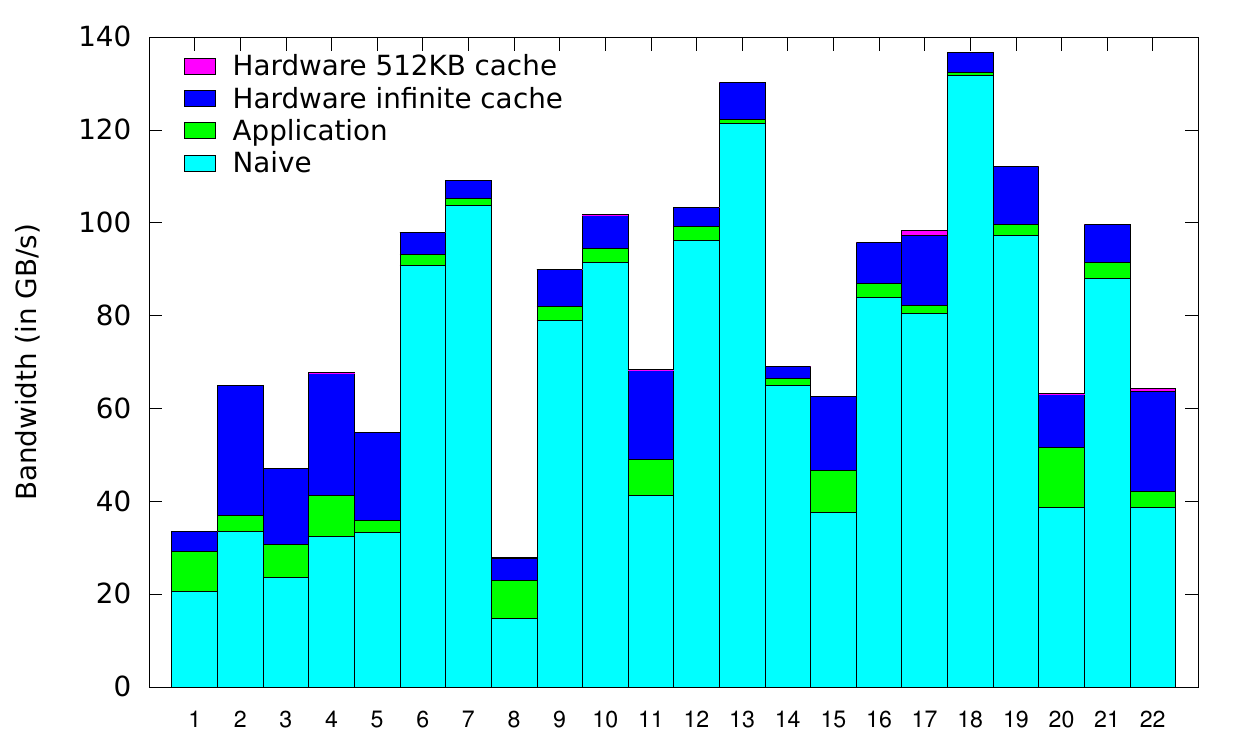}
\caption{Bandwidth achieved on SpMV with different assumptions.}
\label{fig:spmv-bw}
\end{figure}

Figure~\ref{fig:spmv-bw} shows the bandwidth achieved in our
experiments. It shows both naive and actual application
bandwidths. Clearly, the naive approach ignores a significant portion
of the data for some matrices. The application bandwidth obtained
ranges from 22GB/s to 132GB/s. Most matrices have an application
bandwidth below 100GB/s.\looseness=-1

The application bandwidth is computed assuming that every single byte
of the problem is transfered exactly once. This assumption is~(mostly)
true for the matrix and the output vector. However, it does not hold
for the input vector for two reasons: first, it is unlikely that each
element of the input vector will be used by the threads of a single
core, some element will be transferred to multiple cores. Furthermore,
the cache of a Xeon Phi core is only 512kB. The input vector does not
fit in the cache for most matrices and some matrix elements may need
to be transfered multiple times to the same core. We analytically
computed the number of cachelines accessed by each core assuming that
chunks of 64 rows are distributed in a round-robin fashion~(a
reasonable approximation of the dynamic scheduling policy). We
performed the analysis with an infinite cache and with a 512kB
cache. We computed the effective memory bandwidth of SpMV under both
assumptions and display them as the top two stacks of the bars in
Figure~\ref{fig:spmv-bw}. Three observations are striking: first, the
difference between the application bandwidth and estimated actual
bandwidth is greater than 10GB/s on 10 instances and more than 20GB/s
on three of them. The highest difference is seen on {\it
2cubes\_sphere}~(\#2) where the amount of data transferred is 1.7
times larger than the application bandwidth. Second, there is no
difference between the assumed infinite cache and 512kB cache
bandwidth. There is only a visible difference in only two instances:
{\it F1}~(\#17) and {\it cage14}~(\#22). That is, no cache thrashing
occurs. Finally, even when we take the actual memory transfers into
account, the obtained bandwidth is still way below the architecture
peak bandwidth.\looseness=-1

\begin{figure}
\subfigure[{\it pre2}]{\includegraphics[width=.96\linewidth,page=2]{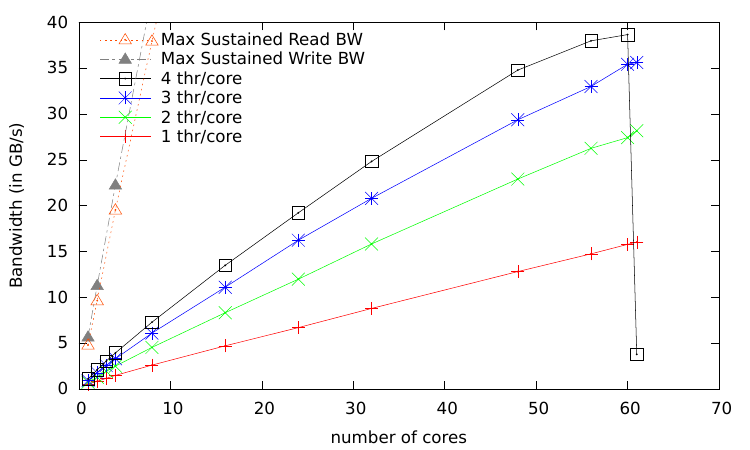}\label{fig:spmv-latencybound}}
\subfigure[{\it crankseg\_2}]{\includegraphics[width=.96\linewidth,page=3]{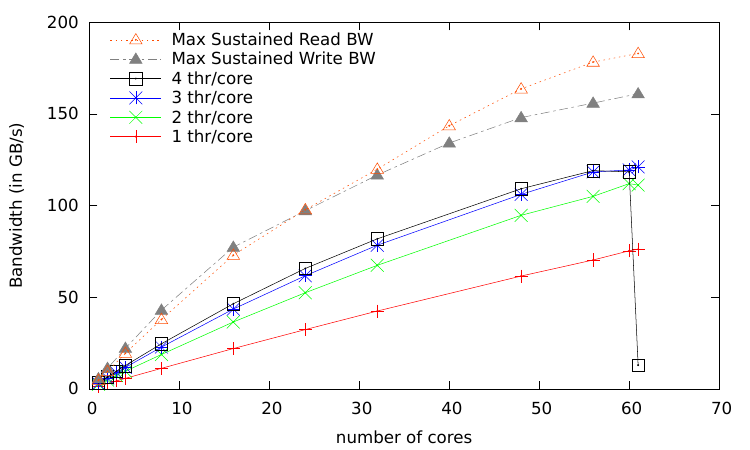}\label{fig:spmv-memorybound}}
\caption{Strong scaling of application bandwidth for two instances~(scheduled with dynamic,64).}
\label{fig:spmv-scaling}
\end{figure}

To understand the bottleneck of this application in a better way, we present
the application bandwidth of two of the instances for various number
of cores and threads per core in Figure~\ref{fig:spmv-scaling}. We
selected these two instances because they seem to be representative
for all the matrices. Most of the instances have a profile similar to
the one presented in Figure~\ref{fig:spmv-latencybound}: there is a
significant difference between the performance achieved with 1, 2, 3,
and 4 threads. Only 5 matrices, {\it crankseg\_2}~(\#13), {\it
pdbHYS}~(\#7), {\it webbase-1M}~(\#8), {\it nd24k}~(\#18), and {\it
torso1}~(\#14), have a profile similar to
Figure~\ref{fig:spmv-memorybound}: there is barely no difference
between the performance achieved with 3 and 4 threads.\looseness=-1

Having that many matrices which show a significant difference between 3 and 4
thread/core performance indicates that there is no saturation either
in the instruction pipeline or in the memory subsystem bandwidth. Most
likely these instances are bounded by the latency of the memory
accesses. As mentioned above, SpMV perform equally well for only 5
instances with 3 and 4 threads per core. These instances are certainly
bounded by an on-core bottleneck: either the core peak memory
bandwidth or the instruction pipeline is full. Three of the instances,
{\it crankseg\_2}, {\it pdbHYS}, and {\it nd24k}, achieves an
application bandwidth over 100GB/s and they are certainly memory
bandwidth bound. However two of them, {\it webbase-1M} and {\it
torso1}, reach a bandwidth lesser than 80GB/s.\looseness=-1

In both figures, the performance obtained when using 61 cores and 4
threads per core is significantly lower. We believe that using all the
threads available in the system hinders some system
operations. However, it is strange that this effect was not visible in
the bandwidth experiments we conducted in Section~\ref{sec:read-bw}
and~\ref{sec:write-bw}.\looseness=-1

\subsection{A First summary}

We summarize what we learned on SpMV with Intel Xeon Phi: maximizing the usefulness of the vgatherd instruction is
the key to optimize the performance: matrices with nonzeros on a row
within the same cacheline (within the same 8-column block in double
precision) will get a higher performance. This is similar to coalesced
memory accesses which is crucial for a good performance while using
GPUs.\looseness=-1

The difference between the application bandwidth and actual used
bandwidth can be high because the same parts of the input vector can
be accessed by different cores. Today, cutting edge processors usually
have 4-8 cores sharing an L3 cache. However, the Intel Xeon Phi coprocessor has 61 of
them, and hence 61 physically different caches. Thus, an improperly
structured sparse matrix could require the same part of the input
vector to be transferred 61 times. This issue is similar to the
problem of input vector distribution in distributed memory
SpMV. Moreover, despite the L2 cache of each core is small by today's
standards, we could not observe a cache thrashing during our
experiments. At last, for most instances, the SpMV kernel appears to
be memory latency bound rather than memory bandwidth bound.\looseness=-1

\subsection{Effect of matrix ordering}

\begin{figure*}[htb]
\centering
\subfigure[Performance]{\includegraphics[width=.54\linewidth,page=1]{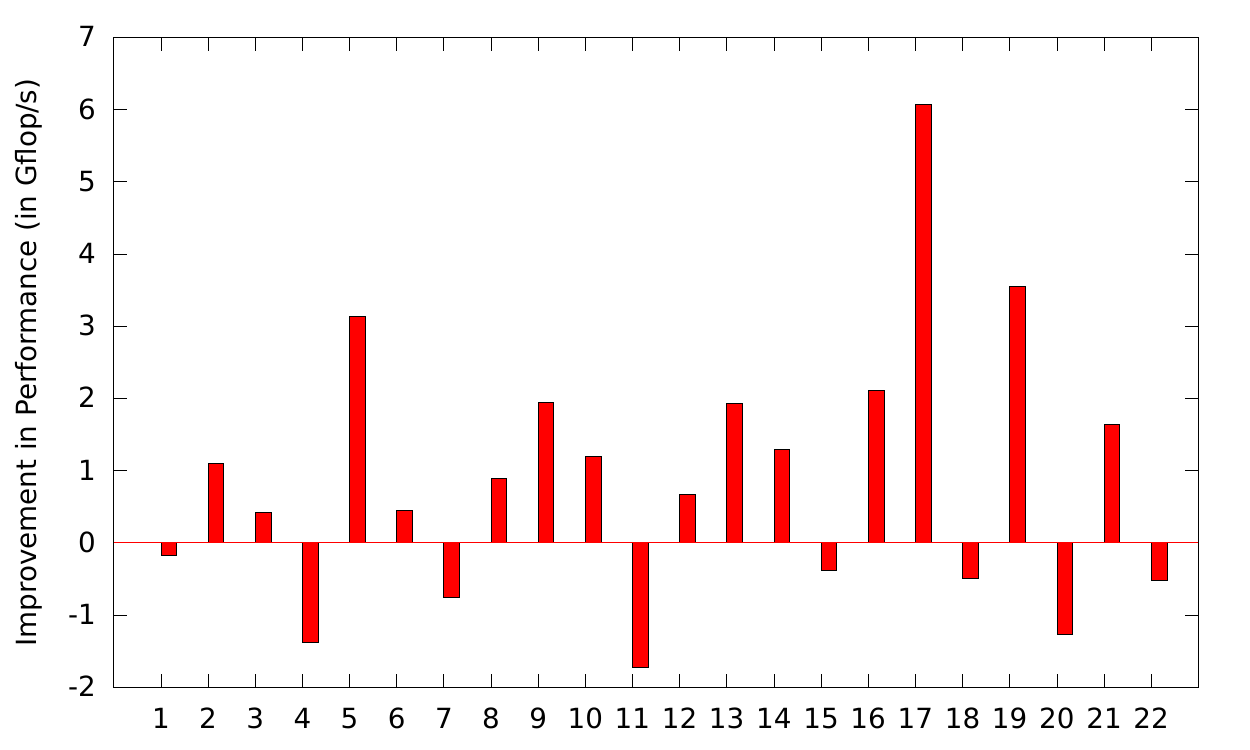}\label{fig:impr-rcm-perf}}
\subfigure[UCLD]{\includegraphics[width=.54\linewidth,page=2]{figures/rcm_analysis.pdf}\label{fig:impr-rcm-ucld}}
\subfigure[Vector
Access]{\includegraphics[width=.54\linewidth,page=3]{figures/rcm_analysis.pdf}\label{fig:impr-rcm-va}}
\caption{Differences on the
      performance and metrics when using RCM ordering. A positive value implies an
  improvement.\looseness=-1}
\label{fig:impr-rcm}
\end{figure*}

A widely-used approach to improve the performance of the SpMV
operation is ordering the rows/columns of the input matrix to make it
more suitable for the kernel. Permuting a set of row and columns is a
technique used in sparse linear algebra for multiple purposes:
improving numerical stability, preconditioning the matrix to reduce
the number of iteration and improving performance. Here, we employ the
MATLAB implementation of the reverse Cuthill-McKee algorithm
(RCM)~\cite{Cuthill69}. RCM has been widely used for minimizing the
maximum distance between the nonzeros and the diagonal of the matrix,
i.e., the {\it bandwidth of the matrix}. It uses a BFS-like vertex/row
order w.r.t. the corresponding graph of the matrix and tries to group
the nonzeros in a region around the diagonal leaving the area away
from the diagonal completely empty. We expect that such a
densification of the nonzeros can improve both the UCLD of the matrix
and reduce the number of times the vector needed to be transfered from
the main memory to the core caches.
The results are presented in Figure~\ref{fig:impr-rcm}. The
improvements in performance, useful cacheline density, and vector
access are given so that a positive value is an improvement and a
negative value is a degradation. The last metric, {\it Vector Access},
represent the expected number of times the input vector needs to be
transferred from the memory.\looseness=-1

The performance improvements are given on
Figure~\ref{fig:impr-rcm-perf}. Matrix {\it F1}~(\#17) obtains the
most absolute improvement which is almost 6GFlop/s. For only 4
matrices, we observe an improvement higher than 2GFlop/s. For most of
the matrices, less than 2GFlop/s improvements in the performance are
observed, and for 8 matrices, we observe a degradation. For 6 of these
matrices, {\it shallow\_water1}~(\#1), {\it mac\_econ}~(\#4), {\it
pdb1HYS}~(\#7), {\it pre2}~(\#11), {\it nd24k}~(\#18), and {\it
cage14}~(\#22), an increase can also be observed on the number of
times the input vector is transferred. Hence, RCM made these matrices
less friendly for the architecture and Vector Access is indeed a
crucial and correlated metric for the performance of SpMV on Xeon Phi.\looseness=-1

\subsection{Effect of register blocking}

One of the limitations in the original SpMV implementation is that
only a single nonzero is processed at a time. The idea of register
blocking is to process all the nonzeros within a region at once. The
region should be small enough that the data associated with it can be
stored in the registers so as to minimize memory
accesses~\cite{Im01}. To implement register blocking, we regularly
partitioned $\Am$ to blocks of size $a \times b$. If $\Am$ has a
nonzero within a region the corresponding block is represented as
dense~(explicitly including the zeros). Hence, the matrix reduces to a
list of non-empty blocks and we represent this list via CRS. Since the
Xeon Phi architecture has a natural alignment on 512 bits, we chose a
block dimension~($a$ or $b$) to be $8$~(8 doubles takes 512 bits). The
other dimension varies from 1 to 8. Each dense block can be
represented using either a row-major order or a column-major order,
depending which dimension is of size 8. When $8 \times 8$ are used,
experiments show that there is no significant difference between the
two schemes.\looseness=-1

To perform the multiplication, each block is loaded~(or streamed) into
the registers in packs of 8 values~(zeros and nonzeros) allowing the
multiplication to be performed using Fused Multiply-Add
operations. Register blocking helps reducing the number of operations
needed to process nonzeros inside a block. Furthermore, it can also
reduce the size of the matrix: if the matrix has $64$ nonzeros in a
dense $8 \times 8$ region it would take $768$ bytes in memory with
CRS~($64 \times 8$ bytes for the value and $64 \times 4$ bytes for the
offset). With register blocking that region is represented in only 516
bytes since only a single offset is required.\looseness=-1

\begin{table}

\centering 
\begin{tabular}{|l||r|r|r|r|r|r|r|}\hline
\bf Configuration        & 8x8 & 8x4 & 8x2 & 8x1 & 4x8 & 2x8 & 1x8 \\\hline
\bf Relative performance  & .53 & .67 & .78 & .92 & .65 & .67 & .64 \\\hline
\bf \# instances improved & 0   &   2 &   5 &   8 &   2 &   1 &   1\\\hline
\end{tabular}
\caption{Performance of register-blocking-based
implementation relative to the one without register blocking. The
second row presents the geometric means of the relative performance of
22 instances: the ratio of Gflop/s obtained with register
blocking to that of the original implementation.\looseness=-1}
\label{tab:reg_block}
\end{table}

The results obtained with register blocking are summarized in
Table~\ref{tab:reg_block}. We omit the matrix transformation times and
only use the timings of matrix multiplications. Overall, we could not
observe an improvement. Partitioning with $8 \times 1$ blocks is the
best scheme and improved performance on 8 instances compared
to the original implementation. When register blocking improved
performance it was never by more than $25\%$. On average, none of the
block sizes helps.\looseness=-1

When large blocks are used, the kernel is memory bound~(the effective
hardware memory bandwidth reaches over 160GB/s in many matrices with
only 3 threads per core). However the average relative performance is
worse since the matrices have low locality. In $8 \times 8$
configuration, less than $35\%$ of the stored values are nonzeros for
most of the matrices. That is less than $\frac{23}{64}$ of the
transferred memory is useful and too much memory is wasted. According
to our experiments, register blocking only saves memory if 70\% of the
values in the blocks are nonzeros. Unfortunately, none of the matrices
respect that condition with $8 \times 8$ blocks. However, when the
blocks are smaller, the density increases and this explains the
increase in relative performance. 10 matrices have more than 50\%
density at size $8 \times 1$~(and 2 have more than 70\% density).\looseness=-1

Our experiments show that register blocking with dense block storage
is not very promising for SpMV on Intel Xeon Phi. Other variants can
be still useful: a logical and straightforward solution is storing the
the blocks via a sparse storage scheme and generate the dense
representation on-the-fly. A 64bit bitmap value would be sufficient to
represent the nonzero pattern in a block~\cite{Buluc11}. Such
representations would improve memory usage and increase the relative
performance. However, they may not be sufficient to surpass the
original implementation two reasons: if the matrices are really sparse
any form of register blocking will provide only a little
improvement. Besides, register blocking does not change the access
pattern to the input vector whose accesses are the one inducing the
latency in the kernel.\looseness=-1

\section{SpMM on Intel Xeon Phi}\label{sec:spmm}

The flop-to-byte ratio of SpMV limits the achievable performance to at
most 30GFlop/s. One simple idea to achieve more performance out of the
Xeon Phi coprocessor is to increase the flop-to-byte ratio by
performing more than one SpMV at a time. In other words, multiplying
multiple vectors will allow us to reach a higher performance. Although
not all the applications can take the advantage of the multiple vector
multiplication at a time, some applications such as graph based
recommendation systems~\cite{Kucuktunc12-ASONAM} or eigensolvers~(by
the use of the LOBPCG algorithm)~\cite{Zhou12-Cluster}
can. Multiplying several vectors by the same matrix boils down to
multiplying a sparse matrix by a dense matrix, which we refer to as
SpMM. All the statements above are also valid for existing
cutting-edge processors and accelerators. However, with its large SIMD
registers, Xeon Phi is expected to perform significantly better.\looseness=-1

In our SpMM implementation for the operation $\Ym \gets \Am\Xm$, the
dense $m \times k$ input matrix $\Xm$ is encoded in row-major, so each
row is contiguous in memory. To process a row $\Am_{i*}$ of the sparse
matrix, a temporary array of size $k$ is first initialized to
zero. Then for each nonzero in $\Am_{i*}$, a row $\Xm_{j*}$ is
streamed to be multiplied by the nonzero and the result is accumulated
into the temporary array. We developed three variants of that
algorithm for Xeon Phi: the first variant is a generic code which
relies on compiler vectorization. The second is tuned for values of
$k$ which are multiple of $8$. This code is manually vectorized to
load the row of the vector by blocks of 8 doubles in a SIMD register
and perform the multiplication and accumulation using Fused
Multiply-Add. The temporary values are kept in registers by taking the
advantage of the large number of SIMD registers available on Xeon
Phi. In the third variant, we use Non-Globally Ordered write
instructions with No-Read hint (NRNGO) and that proved to be fastest in our
bandwidth experiments.\looseness=-1

\begin{figure}[t]
\centering
\subfigure[Comparison of three SpMM variants]{\includegraphics[width=.96\linewidth,page=1]{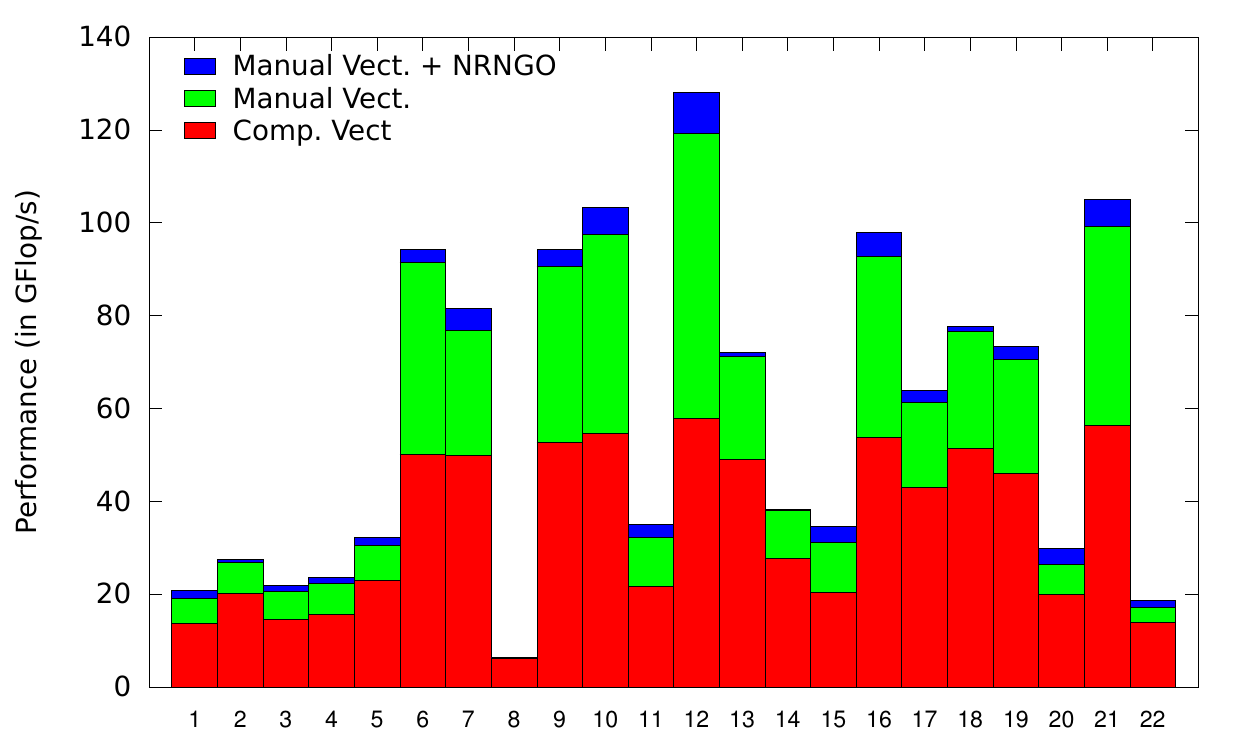}\label{fig:spmm-KNC-GFlops}}
\subfigure[Bandwidth of Manual Vect. + NRNGO]{\includegraphics[width=.96\linewidth,page=3]{figures/KNC-spmm-vect.pdf}\label{fig:spmm-KNC-BW}}
\caption{Performance of SpMM with $k=16$. NRNGO is the variant with
  No-Read hint and Non-Globally ordering.}
\label{fig:spmm-KNC}
\end{figure}

We experimented with $k=16$ and present the results in
Figure~\ref{fig:spmm-KNC}. In many instances, manual vectorization
doubles the performance allowing to reach more than 60GFlop/s in 11
instances. The use of the NRNGO write instructions provides
significant performance improvements. The achieved performance peaks
on the matrix {\it pwtk} matrix at
128GFlop/s. Figure~\ref{fig:spmm-KNC-BW} shows the bandwidth achieved
by the best implementation. The application bandwidth surpasses 60GB/s
in only $1$ instance.  The size of the data is computed as $8 \times m
\times k + 8 \times n \times k + (n+1) \times 4+\tau \times (8+4)$.
However, since there are 16 input vectors, the overhead induced by
transferring the values in $\Xm$ to multiple cores is much higher when
the amount of data transfers are taken into account while computing the
application bandwidth. Once again, the impact of having a finite cache
is only very small. Hence, similar to SpMV, the cache size is not the
problem for Xeon Phi, but having 61 different caches can be a problem
for some applications. \looseness=-1

\section{Against other architectures}\label{sec:comp}

We compare the sparse matrix multiplication performance of Xeon Phi
with 4 other architectures including 2 GPU configurations and 2 CPU
configurations. We used two CUDA-enabled cards from NVIDIA. The NVIDIA
Tesla C2050 is equipped with 448 CUDA Cores clocked 1.15GHz and 2.6GB
of memory clocked at 1.5GHz~(ECC on). This machine uses CUDA driver
and CUDA runtime 4.2. The Tesla K20 comes with 2,496 CUDA Cores
clocked at 0.71GHz and 4.8GB of memory clocked at 2.6GHz~(ECC
on). This machine uses CUDA driver and CUDA runtime 5.0. For both GPU
configurations, we use the CuSparse library as implementation.  We
also use two Intel CPU systems: the first one is a dual Intel Xeon
X5680 configuration, which we call as Westmere. Each processor is
equipped with $6$ cores clocked at $3.33$Ghz and hyperthreading is
disabled. Each processor is equipped with a shared $12$MB L3 cache,
and each core has a local $256$kB L2 cache. The second configuration
is a dual Intel Xeon E5-2670, which we denote with Sandy. Each
processor is equipped with 8 cores clocked at 2.6GHz and
hyperthreading is enabled. Each processor is equipped with a shared
$20$MB L3 cache, and each core has a local $256$kB L2 cache. All the
codes for both CPU architectures are compiled with the {\tt icc 13.0}
with {\tt -O3} optimization flag. All kernels are implemented with
OpenMP. For the performance and stability, all runs are performed with
thread pinning using KMP\_AFFINITY. The implementation used is the
same as the one used on Xeon Phi except the vector optimizations in
SpMM where the instruction sets differ.\looseness=-1

\begin{figure}[htb]
\centering
\subfigure[SpMV]{\includegraphics[width=.99\linewidth]{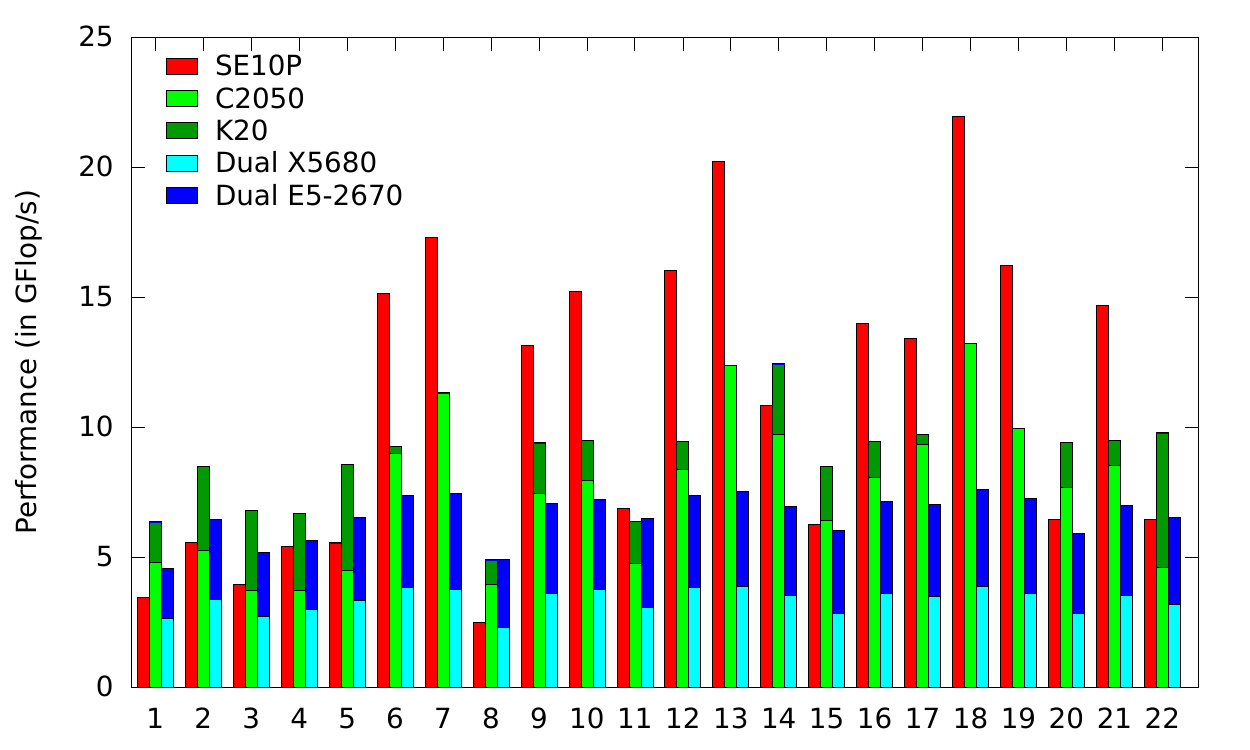}\label{fig:arch-comp-spmv}}
\subfigure[SpMM]{\includegraphics[width=.99\linewidth]{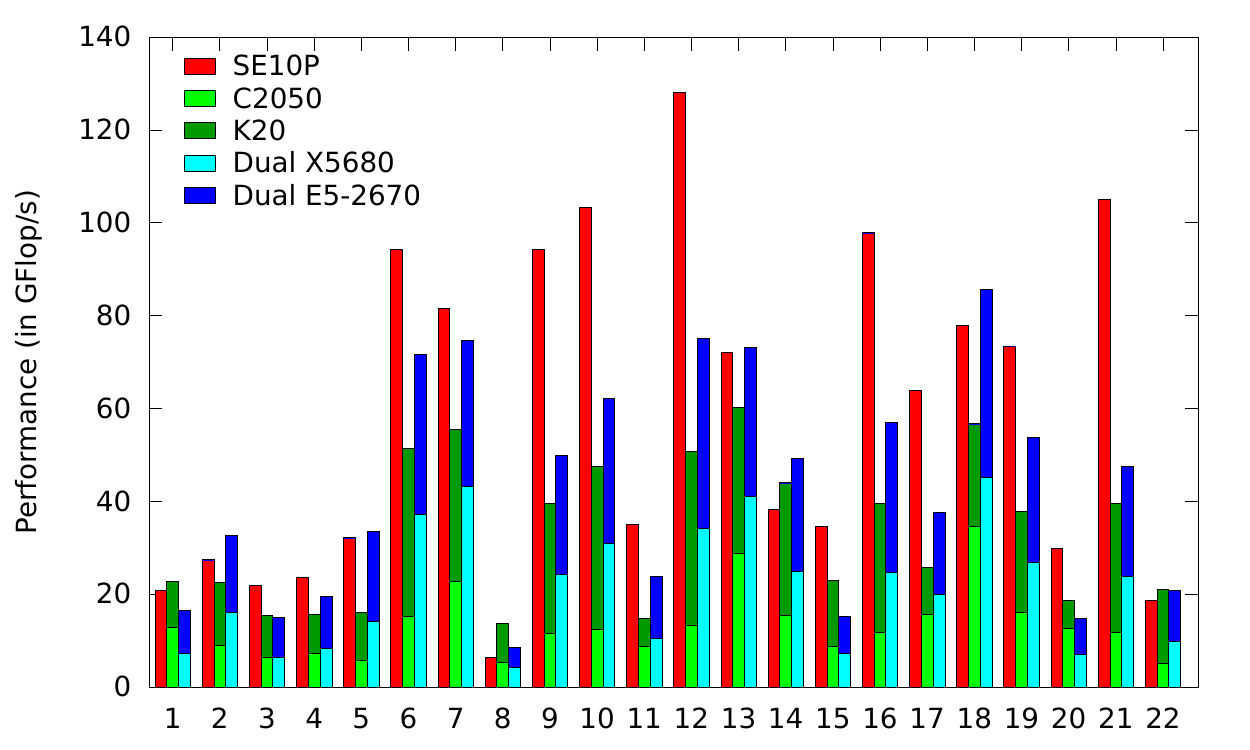}\label{fig:arch-comp-spmm}}
\vspace*{-1em}\caption{Architectural comparison between a Intel Xeon Phi coprocessor
  (Pre-release SE10P), two NVIDIA GPUs (C2050 and K20) and two dual
  CPU architectures (Intel Xeon X5680 and Intel Xeon E5-2670).}
\label{fig:arch-comp}
\end{figure}

Results of the experiments are presented in
Figure~\ref{fig:arch-comp}. Since we have two GPU and two CPU
architectures where one is expected to be better than the
other, we present them as stacked bar charts: K20 on top of C2050
and Sandy on top of Westmere. Figure~\ref{fig:arch-comp-spmv} shows
the SpMV results: Sandy appears to be roughly twice faster than
Westmere. It reaches a performance between 4.5 and 7.6GFlop/s and
achieves the highest performance for one
instance. For GPU architectures, the K20 card is typically faster than
the C2050 card. It performs better for 18 of the 22
instances. It obtains between 4.9 and 13.2GFlop/s and the highest
performance on 9 of the instances. Xeon Phi reaches the highest
performance on 12 of the instances and it is the only architecture
which can obtain more than 15GFlop/s. Furthermore, it does it for 7 of
the instances.\looseness=-1

Figure~\ref{fig:arch-comp-spmm} shows the SpMM results: Sandy gets
twice the performance of the Westmere, which is similar to their
relative SpMV performances. The K20 GPU is often more than twice
faster than C2050, which is much better compared with their relative
performances in SpMV. The Xeon Phi coprocessor gets the best
performance in 14 instances where this number is $5$ and $3$ for the
CPU and GPU configurations, respectively. Intel Xeon Phi is the only
architecture which achieves more than 100GFlop/s. Furthermore, it
reaches more than 60GFlop/s on 9 instances. The CPU configurations
reach more than 60GFlop/s on 6 instances while the GPU configurations
never achieve that performance.\looseness=-1

\section{Conclusion and Future Work}\label{sec:con}

In this work, we analyze the performance of Intel Xeon Phi
coprocessor on SpMV and SpMM. These sparse algebra kernels have been used in many
important applications. To the best of our knowledge, the analysis
gives the first absolute performance results of Intel Xeon Phi.\looseness=-1
  
We showed that the sparse matrix kernels we investigated are latency
bound. Our experiments suggested that having a relatively small 512kB
L2 cache per core is not a problem for Intel Xeon Phi. However, having
61 cores brings increase/decrease in the usefulness of existing
optimization approaches in the literature. Although it is usually a
desired property, having a large number of cores can have a negative
impact on the performance especially if the input data is large and it
is required to be transferred to multiple caches. This increase the
importance of matrix storage schemes, intra-core locality, and data
partitioning among cores. As a future work, we are planning to
investigate such techniques in detail to improve the performance
of Xeon Phi.\looseness=-1

Overall, the performance of the coprocessor is very promising. When
compared with cutting-edge processors and accelerators, its SpMV, and
especially SpMM, performance are superior thanks to its wide registers
and vectorization capabilities. We believe that Xeon Phi will gain
more interest in HPC community in the near future.\looseness=-1

\section*{Acknowledgments}

This work was partially supported by the NSF grants CNS-0643969, OCI-0904809 
and OCI-0904802.\looseness=-1

We would like to thank NVIDIA for the K20 cards, Intel for the Xeon
Phi prototype, and the Ohio Supercomputing Center for access to Intel
hardware. We also would like to thank to Timothy~C.~Prince from Intel
for his suggestions and valuable discussions.\looseness=-1

\begin{small}
\renewcommand{\baselinestretch}{0.88}
\vspace*{3ex} 
\bibliographystyle{abbrv}
\bibliography{paper}

\begin{thebibliography}{10}

\bibitem{Bell09}
N.~Bell and M.~Garland.
\newblock Implementing sparse matrix-vector multiplication on
  throughput-oriented processors.
\newblock In {\em Proc. High Performance Computing Networking, Storage and
  Analysis}, SC '09, pages 18:1--18:11, 2009.

\bibitem{Buluc2009_SPAA}
A.~Bulu\c{c}, J.~T. Fineman, M.~Frigo, J.~R. Gilbert, and C.~E. Leiserson.
\newblock Parallel sparse matrix-vector and matrix-transpose-vector
  multiplication using compressed sparse blocks.
\newblock In {\em Proc. SPAA '09}, pages 233--244, 2009.

\bibitem{Buluc11}
A.~Bulu\c{c}, S.~Williams, L.~Oliker, and J.~Demmel.
\newblock Reduced-bandwidth multithreaded algorithms for sparse matrix-vector
  multiplication.
\newblock In {\em Proc. IPDPS}, 2011.

\bibitem{cramer2012openmp}
T.~Cramer, D.~Schmidl, M.~Klemm, and D.~an~Mey.
\newblock Openmp programming on intel xeon phi coprocessors: An early
  performance comparison.
\newblock In {\em Proceedings of the Many-core Applications Research Community
  (MARC) Symposium at RWTH Aachen University}, Nov. 2012.

\bibitem{Cuthill69}
E.~Cuthill and J.~McKee.
\newblock Reducing the bandwidth of sparse symmetric matrices.
\newblock In {\em Proc. ACM national conference}, pages 157--172, 1969.

\bibitem{Eisenlohr12-TACC}
J.~Eisenlor, D.~E. Hudak, K.~Tomko, and T.~C. Prince.
\newblock Dense linear algebra factorization in {OpenMP} and {Cilk Plus on
  Intel MIC}: Development experiences and performance analysis.
\newblock In {\em TACC-Intel Highly Parallel Computing Symp.}, 2012.

\bibitem{Im01}
E.-J. Im and K.~A. Yelick.
\newblock Optimizing sparse matrix computations for register reuse in sparsity.
\newblock In {\em Proc. of ICCS}, pages 127--136, 2001.

\bibitem{Jain08}
A.~Jain.
\newblock {pOSKI}: An extensible autotuning framework to perform optimized
  spmvs on multicore architecture.
\newblock Master's thesis, UC Berkeley, 2008.

\bibitem{Krotkiewski10}
M.~Krotkiewski and M.~Dabrowski.
\newblock Parallel symmetric sparse matrix-vector product on scalar multi-core
  {CPU}s.
\newblock {\em Parallel Comput.}, 36(4):181--198, Apr. 2010.

\bibitem{Kucuktunc12-ASONAM}
O.~K{\"u}{\c{c}}{\"u}ktun{\c{c}}, K.~Kaya, E.~Saule, and {\"U}.~V.
  {\c{C}}ataly{\"u}rek.
\newblock Fast recommendation on bibliographic networks.
\newblock In {\em Proc. ASONAM'12}, Aug 2012.

\bibitem{Mellor-Crummey04}
J.~Mellor-Crummey and J.~Garvin.
\newblock Optimizing sparse matrix-vector product computations using unroll and
  jam.
\newblock {\em Int. J. High Perform. Comput. Appl.}, 18(2), May 2004.

\bibitem{Nishtala07}
R.~Nishtala, R.~W. Vuduc, J.~W. Demmel, and K.~A. Yelick.
\newblock When cache blocking of sparse matrix vector multiply works and why.
\newblock {\em Appl. Algebra Eng., Commun. Comput.}, 18(3):297--311, May 2007.

\bibitem{Potluri12-TACC}
S.~Potluri, K.~Tomko, D.~Bureddy, and D.~K. Panda.
\newblock {Intra-MIC MPI} communication using {MVAPICH2}: Early experience.
\newblock In {\em TACC-Intel Highly Parallel Computing Symp.}, 2012.

\bibitem{Saad94sparskit}
Y.~Saad.
\newblock Sparskit: a basic tool kit for sparse matrix computations - version
  2, 1994.

\bibitem{Saule12-MTAAP}
E.~Saule and {\"U}.~V. {\c{C}}ataly{\"u}rek.
\newblock An early evaluation of the scalability of graph algorithms on the
  {Intel MIC} architecture.
\newblock In {\em IPDPS Workshop MTAAP}, 2012.

\bibitem{Stock12-TACC}
K.~Stock, L.-N. Pouchet, and P.~Sadayappan.
\newblock Automatic transformations for effective parallel execution on intel
  many integrated core.
\newblock In {\em TACC-Intel Highly Parallel Computing Symp.}, 2012.

\bibitem{Vuduc05}
R.~Vuduc, J.~Demmel, , and K.~Yelic.
\newblock {OSKI}: A library of automatically tuned sparse matrix kernels.
\newblock In {\em Proc. SciDAC 2005, J. of Physics: Conference Series}, 2005.

\bibitem{Williams07}
S.~Williams, L.~Oliker, R.~Vuduc, J.~Shalf, K.~Yelick, and J.~Demmel.
\newblock Optimization of sparse matrix-vector multiplication on emerging
  multicore platforms.
\newblock In {\em Proc. SC '07}, pages 38:1--38:12, 2007.

\bibitem{Zhou12-Cluster}
Z.~Zhou, E.~Saule, H.~M. Aktulga, C.~Yang, E.~G. Ng, P.~Maris, J.~P. Vary, and
  {\"U}.~V. {\c{C}}ataly{\"u}rek.
\newblock An out-of-core eigensolver on {SSD}-equipped clusters.
\newblock In {\em Proc. of IEEE Cluster}, Sep 2012.

\end{thebibliography}
\renewcommand{\baselinestretch}{1}
\end{small} 
\end{document}